\documentclass[preprint,aps,nofootinbib]{revtex4}
\input{epsf.tex}
\usepackage[dvips]{graphicx,epsfig}
\usepackage{subfigure}
\usepackage{textcomp}
\def\bkR{{\rm I\kern-.17em R}}
\def\bkC{{\rm \kern.24em \vrule width.05em height1.4ex depth-.05ex \kern-.26em C}}

\def\be{\beta}

\def\frac#1#2{{\textstyle{{#1}\over {#2}}}}
\def\lsim{\mathrel{\rlap{\lower4pt\hbox{\hskip1pt$\sim$}}
    \raise1pt\hbox{$<$}}}
\def\gsim{\mathrel{\rlap{\lower4pt\hbox{\hskip1pt$\sim$}}
    \raise1pt\hbox{$>$}}}
\def\sqr#1#2{{\vcenter{\vbox{\hrule height.#2pt
         \hbox{\vrule width.#2pt height#1pt \kern#1pt
         \vrule width.#2pt}
         \hrule height.#2pt}}}}

\def\laq{\raise 0.4 ex \hbox{$<$}\kern -0.8 em\lower 0.62 ex\hbox{$\sim$}}
\def\gaq{\raise 0.4 ex \hbox{$>$}\kern -0.7 em\lower 0.62 ex\hbox{$\sim$}}

\def\be{\begin{equation}}
\def\ee{\end{equation}}
\def\ba{\begin{eqnarray}}
\def\ea{\end{eqnarray}}

\def\dalemb#1#2{{\vbox{\hrule height.#2pt
        \hbox{\vrule width.#2pt height#1pt \kern#1pt \vrule width.#2pt}
        \hrule height.#2pt}}}

\def\dalemb#1#2{{\vbox{\hrule height.#2pt
        \hbox{\vrule width.#2pt height#1pt \kern#1pt \vrule width.#2pt}
        \hrule height.#2pt}}}

\def\gtorder{\mathrel{\raise.3ex\hbox{$>$}\mkern-14mu
             \lower0.6ex\hbox{$\sim$}}}
\def\ltorder{\mathrel{\raise.3ex\hbox{$<$}\mkern-14mu
             \lower0.6ex\hbox{$\sim$}}}

\begin{document}

\rightline{DF/IST-8.2007}
\rightline{July 2008}

\title{Phase-Space Noncommutative Quantum Cosmology}

\author{Catarina Bastos\footnote{Also at Instituto de Plasmas e Fus\~ao Nuclear,
IST. cbastos@fisica.ist.utl.pt},
Orfeu Bertolami\footnote{Also at Instituto de Plasmas e Fus\~ao Nuclear,
IST. orfeu@cosmos.ist.utl.pt}}

\vskip 0.3cm

\affiliation{Departamento de F\'\i sica, Instituto Superior T\'ecnico \\
Avenida Rovisco Pais 1, 1049-001 Lisboa, Portugal}

\author{Nuno Costa Dias\footnote{Also at Grupo de F\'{\i}sica Matem\'atica, UL,
Avenida Prof. Gama Pinto 2, 1649-003, Lisboa, Portugal. ncdias@mail.telepac.pt}, Jo\~ao Nuno Prata\footnote{Also at Grupo de F\'{\i}sica Matem\'atica, UL,
Avenida Prof. Gama Pinto 2, 1649-003, Lisboa, Portugal. joao.prata@mail.telepac.pt}}

\vskip 0.3cm

\affiliation{Departamento de Matem\'{a}tica, Universidade Lus\'ofona de
Humanidades e Tecnologias \\
Avenida Campo Grande, 376, 1749-024 Lisboa, Portugal}

%\date{}

\vskip 0.5cm

\begin{abstract}

\vskip 1cm

{We present a phase-space noncommutative extension of Quantum Cosmology and study the Kantowski-Sachs (KS) cosmological model requiring that the two scale factors of the KS metric, the coordinates of the system, and their conjugate canonical momenta do not commute. Through the Arnowitt-Deser-Misner formalism, we obtain the Wheeler-DeWitt (WDW) equation for the noncommutative system. The Seiberg-Witten map is used to transform the noncommutative equation into a commutative one, i.e. into an equation with commutative variables, which depend on the noncommutative parameters, $\theta$ and $\eta$. Numerical solutions are found both for the classical and the quantum formulations of the system. These solutions are used to characterize the dynamics and the state of the universe. From the classical solutions we obtain the behavior of quantities such as the volume expansion, the shear and the characteristic volume. However the analysis of these quantities does not lead to any restriction on the value of the noncommutative parameters, $\theta$ and $\eta$. On the other hand, for the quantum system, one can obtain, via the numerical solution of the WDW equation, the wave function of the universe both for commutative as well as for the noncommutative models. Interestingly, we find that the existence of suitable solutions of the WDW equation imposes bounds on the values of the noncommutative parameters. Moreover, the noncommutativity in the momenta leads to damping of the wave function implying that this noncommutativity can be of relevance for the selection of possible initial states of the early universe.}

\end{abstract}

\maketitle

\section{Introduction}

Noncommutative space-time and its physical implications have recently been studied with great interest. This interest has its roots in developments in String Theory/M-Theory, where a noncommutative effective low-energy gauge theory action naturally arises when one describes the low energy excitations of open strings in the presence of a Neveu-Schwarz constant background $B$ field \cite{Connes,Seiberg}. Noncommutative theories are also considered to explain some physical effects such as the Quantum Hall effect \cite{Belissard} and the noncommutative Landau problem \cite{Gamboa,Horvathy}. Moreover, noncommutative extensions of the gravitational quantum well have also been examined in connection with the measurement of the first two quantum states of the gravitational quantum well for ultra cold neutrons \cite{Bertolami1,Bertolami2}.

In this work we shall assume that the noncommutativity of space-time is a characteristic feature of quantum gravity and that its effects should be significant at very high energy scales, at the early universe. Thus, it is natural to consider the role of noncommutative geometry in the context of quantum cosmology. Before considering in detail our quantum cosmological setting let us review the main ideas behind the noncommutative extensions of quantum mechanics.

The usual formulations of noncommutative quantum mechanics (NCQM) considered in the literature (\cite{Bertolami1}-\cite{Bastos}) are based on canonical extensions of the Heisenberg algebra. Time is required to be a commutative parameter and the theory lives in a $2d$-dimensional phase-space of operators with noncommuting position and momentum variables. The extended Heisenberg algebra reads:
\be\label{eq1.1}
\left[\hat q_i, \hat q_j \right] = i\theta_{ij} \hspace{0.2 cm}, \hspace{0.2 cm} \left[\hat q_i, \hat p_j \right] = i \hbar \delta_{ij} \hspace{0.2 cm},
\hspace{0.2 cm} \left[\hat p_i, \hat p_j \right] = i \eta_{ij} \hspace{0.2 cm},  \hspace{0.2 cm} i,j= 1, ... ,d
\ee
where $\eta_{ij}$ and $\theta_{ij}$ are antisymmetric real constant ($d \times d$) matrices and $\delta_{ij}$ is the identity matrix. Theoretical predictions for specific noncommutative systems have been compared with experimental data leading to bounds on the noncommutative parameters obtained in the field theory and gravitational quantum well context, respectively  \cite{Carroll,Bertolami1}. At those energy scales, the bounds for the noncommutative parameters are
\be\label{eq1.2}
\theta \leq 4 \times 10^{-40} m^2 \hspace{0.5 cm},\hspace{0.5 cm} \eta \leq 1.76 \times 10^{-61} kg^2 m^2 s^{-2}.
\ee
A great deal of work has been devoted to studying the structural and formal aspects of the quantum theory based on the algebra (\ref{eq1.1}).
The extended Heisenberg algebra is related to the standard Heisenberg algebra:
\be\label{eq1.3}
\left[\hat R_i, \hat R_j \right] = 0 \hspace{0.2 cm}, \hspace{0.2 cm} \left[\hat R_i, \hat \Pi_j \right]
= i \hbar \delta_{ij} \hspace{0.2 cm}, \hspace{0.2 cm} \left[\hat \Pi_i, \hat \Pi_j \right] = 0 \hspace{0.2 cm},
\hspace{0.2 cm} i,j= 1, ... ,d ~,
\ee
by a class of linear (non-canonical) transformations:
\be\label{eq1.4}
\hat q_i = \hat q_i \left(\hat R_j , \hat \Pi_j \right) \hspace{0.2 cm},\hspace{0.2 cm}
\hat p_i = \hat p_i \left(\hat R_j , \hat \Pi_j \right)
\ee
which are often referred to as the Seiberg-Witten (SW) map \cite{Seiberg}. With these transformations, one is able to convert a noncommutative system into a modified commutative system, which is dependent on the noncommutative parameters and of the particular SW map. The states of the system are then wave functions of the ordinary Hilbert space and the dynamics is determined by the usual Schr\"{o}dinger equation with a modified $\eta,\theta$-dependent Hamiltonian. One stresses however, that the physically relevant quantities such as expectation values, probabilities and eigenvalues of operators are independent of the chosen SW map \cite{Bastos}.

In this paper we study a noncommutative extension of Quantum Cosmology (QC). We assume that space-time noncommutativity is significant at very high energy scales and thus that nontrivial effects might have emerged at early times. We use the canonical quantization prescription to obtain a minisuperspace quantum model for the Universe, arising from the WDW equation (see e.g. \cite{Hartle,Bertolami3} and references therein)
based on a KS metric.

The KS cosmological model has been previously studied in the context of noncommutative quantum cosmology, although only for the case where just the configuration variables are noncommutative \cite{Compean,Barbosa}. Here, we shall extend noncommutativity to the momentum sector as well. This provides a more general formulation, which displays several distinctive features. Moreover, it also provides the natural setting where to analyze the influence of the magnitude of the noncommutative parameters on the overall behavior of the theory. Indeed, noncommutativity of the momentum sector should not be discarded, as there are instances where the momentum noncommutative corrections may be larger and more susceptible to experimental detection \cite{Bertolami1}. In this paper we shall study both the classical and the quantum formulations of the full noncommutative KS cosmological model. Phase-space noncommutativity in Quantum Cosmology has been considered previously \cite{Khosravi1,Khosravi2}. However, this has been done in a different context, namely that of multidimensional cosmology.

At the classical level, the effect of noncommutativity can be studied through the behavior of physical quantities such as the volume expansion, $\Theta(t)$, with respect to the proper time of a co-moving observer, the shear, $\sigma(t)$, and a characteristic length scale, $l(t)$ (see e.g. Ref. \cite{Ryan} for an extensive discussion). If from the qualitative point of view, noncommutativity in the configuration variables leads to no major effect when compared with the commutative case, a non-trivial noncommutativity in momenta introduces a distinct effect in what concerns the behavior of the shear (c.f. Figure 1 below). It is relevant to point out however, that the analysis of the classical noncommutative model, either in configuration space or in phase space, does not yield any bound or restriction of the possible values of the noncommutative parameters $\eta$ and $\theta$ and of the relevant canonical conjugate momenta of the KS model.

This picture changes at the quantum level. Here, our approach is tantamount to converting the full noncommutative model into a modified commutative system using a suitable SW map. This yields a deformation of the minisuperspace due to the noncommutativity of the variables. By examining the physical solutions of the WDW equation we find that they exist only for particular values of the noncommutative parameters. This is particularly relevant as the most natural outcome of quantum gravity is likely to involve noncommutative features.

Furthermore, we will see that noncommutativity leads to a richer structure of states for the early universe. However, this is the case only when momenta noncommutativity is included. In this case the fundamental solutions of the WDW equation (which are featureless oscillations for both the commutative and configuration noncommutative cases) display a damping behavior. We expect, by refining the cosmological model and /or by choosing other deformations of the Heisenberg algebra, to obtain normalized solutions of the WDW equation. This will provide a major breakthrough for the physical interpretation of the initial state of the universe.

This paper is organized as follows. In Section 2, we review the original formulation of the commutative classical and quantum KS cosmological model. We extend the formalism to encompass noncommutativity in both coordinates and momenta. We consider in detail the classical and the quantum formulations of this noncommutative extension and obtain the noncommutative WDW equation. In Section 3, we present our numerical solutions for the classical Hamiltonian equations and for the WDW equation. We analyze the classical behavior of three relevant physical quantities, the volume expansion, shear and a characteristic volume. We then numerically solve the WDW equation and depict some typical wave functions analyzing the set of values for $\theta$ and $\eta$ for which a solution exists and the wave function has features such as damping behavior. Finally, in Section 4, we discuss our results and put forward some conclusions.

\section{The cosmological model}

Let us consider a cosmological model given by the KS metric, which has the correspondent line element given by
\be\label{eq2.0}
ds^2=-N^2dt^2+X^2(t)dr^2+Y^2(t)(d\theta^2+\sin^2{\theta}d\varphi^2)~.
\ee
In the Misner parametrization, this can be written as \cite{Ryan}
\be\label{eq2.1}
ds^2=-N^2dt^2+e^{2\sqrt{3}\beta}dr^2+e^{-2\sqrt{3}\beta}e^{-2\sqrt{3}\Omega}(d\theta^2+\sin^2{\theta}d\varphi^2)~,
\ee
where $\beta$ and $\Omega$ are scale factors and $N$ is the lapse function. The presence of at least two scale factors is necessary to consider a noncommutative extension of the classical problem. Following the ADM construction, one can derive the Hamiltonian for this metric,
\be\label{eq2.1a}
H=N{\cal H}=Ne^{\sqrt{3}\beta+2\sqrt{3}\Omega}\left[-{P_{\Omega}^2\over24}+{P_{\beta}^2\over24}-2e^{-2\sqrt{3}\Omega}\right]~,
\ee
where $P_{\Omega}$ and $P_{\beta}$ are the canonical momenta conjugate  to $\Omega$ and $\beta$, respectively. The lapse function, $N$, will be taken to be $N=24e^{-\sqrt{3}\beta-2\sqrt{3}\Omega}$. This corresponds to a particular gauge choice, which is motivated by technical simplicity (related only to the classical treatment). Our results are nevertheless, all gauge independent. At the quantum level, the treatment is manifestly covariant as the lapse function does not enter at all in the formalism. At the classical level, only the dynamics of the fundamental variables is gauge dependent while the three relevant physical variables (the volume expansion, the shear and the characteristic volume) are gauge invariant. We will see this explicitly in the next sections. We will consider the classical and the quantum formulations separately.

\subsection{The Classical Model}

Classically, the equations of motion for the phase-space variables $\Omega$, $\beta$, $P_{\Omega}$ and $P_{\beta}$ can be obtained from the Poisson brackets algebra. For the commutative case, the Poisson brackets are:
\be\label{eq2.3a}
\left\{\Omega,P_{\Omega}\right\}=1\hspace{0.1 cm},\hspace{0.1 cm}\left\{\beta,P_{\beta}\right\}=1\hspace{0.1 cm},\hspace{0.1 cm}\left\{\Omega,\beta\right\}=0\hspace{0.1 cm},\hspace{0.1 cm}\left\{P_{\Omega},P_{\beta}\right\}=0~.
\ee
and thus, the equations of motion with respect to the internal time are the usual Hamiltonian equations $\dot X=N\{X,{\cal H}\}$ for each of the canonical variables $X$. In the constraint hypersurface
\be\label{eq2.3a1}
{\cal H}\approx 0
\ee
this leads to \cite{Barbosa}:
\ba\label{eq2.3b}
&&\dot{\Omega}=-2P_{\Omega}~,\nonumber\\
&&\dot{P_{\Omega}}=-96\sqrt{3}e^{-2\sqrt{3}\Omega}~,\nonumber\\
&&\dot{\beta}=2P_{\beta}~,\nonumber\\
&&\dot{P_{\beta}}=0~.
\ea
The solutions for $\Omega$ and $\beta$ are
\ba\label{eq2.3c}
&&\Omega(t)={\sqrt{3}\over6}\ln{\left({48\over P_{\beta_0}^2}\cosh^2{\left[2\sqrt{3}P_{\beta_0}(t-t_0)\right]}\right)}~,\nonumber\\
&&\beta(t)=2P_{\beta_0}(t-t_0)+\beta_0~.
\ea

In previous works the noncommutative extension of this model has been considered \cite{Compean,Barbosa}. However, this has only been done for spatial noncommutativity (i.e. for noncomutative configuration variables $\Omega$ and $\beta$). In Ref. \cite{Barbosa}, the authors obtained classical solutions for the system described by the Hamiltonian constraint (\ref{eq2.3a1}) in the context of a noncommutative phase-space with symplectic structure given by Eqs. (\ref{eq2.3a}) with $\{\Omega,\beta\}=\theta,$ instead of $\{\Omega,\beta\}=0$.

Clearly, a more general noncommutative extension can be obtained by imposing a noncommutative relation between the two coordinates, $\Omega$ and $\beta$, and also between the two canonical momenta, $P_{\Omega}$ and $P_{\beta}$, as follows:
\be\label{eq2.10}
\left\{\Omega,P_{\Omega}\right\}=1\hspace{0.1 cm},\hspace{0.1 cm}\left\{\beta,P_{\beta}\right\}=1\hspace{0.1 cm},\hspace{0.1 cm}\left\{\Omega,\beta\right\}=\theta\hspace{0.1 cm},\hspace{0.1 cm}\left\{P_{\Omega},P_{\beta}\right\}=\eta~,
\ee
In this case the classical equations of motion for the noncommutative system are
\ba\label{eq2.11}
&&\dot{\Omega}=-2P_{\Omega}~,\hspace{0.5cm}(a)\nonumber\\
&&\dot{P_{\Omega}}=2\eta P_{\beta}-96\sqrt{3}e^{-2\sqrt{3}\Omega}~,\hspace{0.5cm}(b)\nonumber\\
&&\dot{\beta}=2P_{\beta}-96\sqrt{3}\theta e^{-2\sqrt{3}\Omega}~,\hspace{0.5cm}(c)\nonumber\\
&&\dot{P_{\beta}}=2\eta P_{\Omega}~.\hspace{0.5cm}(d)
\ea
It seems that an analytical solution of this system is beyond reach, given the entanglement among the four variables. On
the other hand a numerical solution can be obtained and used to provide predictions for several physical relevant quantities. We will proceed in this way in the next section. But before that, let us point out that
Eqs. (\ref{eq2.11}a) and (\ref{eq2.11}d) yield a constant of motion:
\be\label{eq3.1}
\dot{P_{\beta}}=-\eta(-2P_{\Omega})=-\eta\dot{\Omega}\Rightarrow P_{\beta}+\eta\Omega=C~,
\ee
that will play an important role in solving the noncommutative WDW equation.

\subsection{The Quantum Model}

Here and henceforth, we assume a system of units where $c=\hbar=G=1$. Consequently, the noncommutative parameters,
$\theta$ and $\eta$, being an intrinsic feature of quantum gravity should be of order one since so is the Planck length, $L_P=1$.

The canonical quantization of the classical Hamiltonian constraint Eq. (\ref{eq2.3a1}) yields the commutative WDW equation for the wave function of the universe. For the simplest factor ordering of operators this equation reads
\be\label{eq2.2}
\exp{(\sqrt{3} \hat{\beta}+2\sqrt{3} \hat{\Omega})}\left[- \hat P^2_{\Omega}+ \hat P^2_{\beta}-48e^{-2\sqrt{3} \hat{\Omega}}\right]\psi(\Omega,\beta)=0~.
\ee
where $\hat P_{\Omega}=-i \frac{\partial }{\partial \Omega}$, $\hat P_{\beta}=-i \frac{\partial }{\partial \beta}$ are the fundamental momentum operators conjugate to $\hat{\Omega} = \Omega$ and $\hat{\beta} = \beta$, respectively. Notice that the Eq.(\ref{eq2.2}) is dependent of the prescribed factor order. This is, however, a common feature to all quantum cosmological models (both commutative and noncommutative). Indeed, we may say that the full identification of the quantum cosmological model requires specifying an operator ordering. For our model we choose the simplest factor order, which has already been studied in the past both for the commutative and the configuration noncommutative cases. This allows us to compare our results with the previous ones found in the literature.

The solutions of (\ref{eq2.2}) can be shown to be of the form \cite{Compean}
\be\label{eq2.3}
\psi^{\pm}_{\nu}(\Omega,\beta)=e^{\pm i\nu\sqrt{3}\beta}K_{i\nu}(4e^{-\sqrt{3}\Omega})~,
\ee
where $K_{i\nu}$ are modified Bessel functions.

We now require the coordinates and the canonical momenta to be noncommutative and obtain the extended Heisenberg algebra,
\be\label{eq2.4}
\left[\hat{\Omega}, \hat{\beta} \right]=i\theta\hspace{0.2 cm},\hspace{0.2 cm}\left[\hat P_{\Omega}, \hat P_{\beta}\right]=i\eta\hspace{0.2 cm},\hspace{0.2 cm}\left[\hat{\Omega}, \hat P_{\Omega}\right]=\left[\hat{\beta},\hat P_{\beta}\right]=i~.
\ee
Our strategy to obtain a representation of the algebra (\ref{eq2.4}) is to transform it into the standard Heisenberg algebra through a suitable non-unitary linear transformation, dubbed as SW map:
\ba\label{eq2.8}
\hat{\Omega} =\lambda \hat{\Omega}_{c}-{\theta\over2\lambda} \hat P_{\beta_c} \hspace{0.2cm} , \hspace{0.2cm} \hat{\beta} = \lambda \hat{\beta}_{c} + {\theta\over2\lambda} \hat P_{\Omega_c}~,\nonumber\\
\hat P_{\Omega}= \mu \hat P_{\Omega_c} + {\eta\over2\mu} \hat{\beta}_{c} \hspace{0.2cm} , \hspace{0.2cm} \hat P_{\beta}=\mu \hat P_{\beta_c}- {\eta\over2\mu} \hat{\Omega}_{c}~,
\ea
where the index $c$ denotes commutative variables, i.e. variables for which $\left[\hat{\Omega}_c, \hat{\beta}_c\right]=\left[\hat P_{\Omega_c}, \hat P_{\beta_c}\right]=0$ and $\left[\hat{\Omega}_c, \hat P_{\Omega_c}\right]=\left[\hat{\beta}_c, \hat P_{\beta_c}\right]=i$. This transformation can be inverted, provided:
\be\label{eq3.1a}
\xi \equiv \theta \eta <1.
\ee
In that case the inverse transformation reads:
\ba\label{eq3.2}
\hat{\Omega}_c={1\over\sqrt{1- \xi}}\left( \mu \hat{\Omega} + {\theta\over2\lambda} \hat P_{\beta}\right) \hspace{0.2cm} , \hspace{0.2cm} \hat{\beta}_c={1\over\sqrt{1-\xi}} \left( \mu \hat{\beta} -{\theta\over2\lambda} \hat P_{\Omega}\right)~,\nonumber\\
\hat P_{\Omega_c}={1\over\sqrt{1-\xi}} \left(\lambda \hat P_{\Omega}-{\eta\over2\mu} \hat{\beta} \right) \hspace{0.2cm} , \hspace{0.2cm} \hat P_{\beta_c}={1\over\sqrt{1-\xi}}\left( \lambda \hat P_{\beta}+{\eta\over2\mu} \hat{\Omega} \right)~.
\ea
Substituting the noncommutative variables, expressed in terms of commutative ones, into the commutation relations (\ref{eq2.4}), one obtains a relation between the dimensionless constants $\lambda$ and $\mu$:
\be\label{eq2.8a}
\left(\lambda\mu\right)^2-\lambda\mu+{\xi\over4}=0\Leftrightarrow\lambda\mu={1+\sqrt{1-\xi}\over2}~.
\ee
Using the transformation (\ref{eq2.8}), one may regard (\ref{eq2.4}) as an algebra of operators acting on the usual Hilbert space $L^2(\bkR^2)$. In this representation the WDW equation (\ref{eq2.2}) is deformed into a modified second order partial differential equation, which exhibits an explicit dependence on the noncommutative parameters:
\be\label{eq2.9}
\left[-\left(-i \mu {\partial \over \partial {\Omega_c}}+{\eta\over2\mu}\beta_{c}\right)^2+\left(-i \mu {\partial \over \partial {\beta_c}}-{\eta\over2\mu}\Omega_c\right)^2-48\exp{\left[-2\sqrt{3}\left(\lambda\Omega_c+i{\theta\over2\lambda} {\partial \over \partial {\beta_c}} \right)\right]}\right]\psi(\Omega_c,\beta_c)=0~.
\ee
This equation is fairly complex and cannot be fully solved analytically. However, the noncommutative quantum version of the constant of motion Eq. (\ref{eq3.1}):
\be\label{eq3.3}
\hat{C}=\hat{P_{\beta}}+\eta\hat{\Omega}=\sqrt{1-\xi}\left(\mu \hat{P}_{\beta_c}+{\eta\over2\mu}\hat{\Omega}_c\right)
\ee
commutes with the noncommutative Hamiltonian constraint Eq. (\ref{eq2.9}). We stress once again that this is only valid for the chosen operator ordering. This allows one to transform the partial differential equation (\ref{eq2.9}) into an ordinary differential equation, which can be then solved numerically. We will present these results in the next section.

\section{Solutions}

\subsection{Classical solutions}

The relevant physical variables for the classical system are: the volume expansion, $\Theta(t)$, with respect to a time-like vector field U, which we parametrize by proper time, so that $U\cdot U=-1$; the shear, $\sigma(t)=[\sigma^{\mu\tau}\sigma_{\mu\tau}]^{1/2}$, where $\sigma_{\mu\nu}=[(1/2)(u_{\rho;\tau}+u_{\tau;\rho})-(1/3)\Theta h_{\rho\tau}]h^{\rho}_{\mu}h^{\tau}_{\nu}$ is the shear tensor \cite{Ryan}; and a characteristic volume for the metric, $l^3(t)$. In these definitions the semi-colon stands for covariant derivative. $h_{\mu\nu}$ are the components of the tensor $h$,which is the projection onto the set of vectors perpendicular to $U$, and $u_{\mu}$ are the covariant components of $U$ in four dimensions, given by $u_{\mu}=g_{\mu\rho}u^{\rho}$. The volume expansion and the shear are related with the dynamics of the space-time and measure the rate at which an element of volume in the universe deforms \cite{Ryan}. The characteristic volume can be obtained from a characteristic length scale, $l(t)$, which is defined in terms of the volume expansion as $\Theta=3\dot{l}/(lN)$ \cite{Barbosa}. For the KS metric these quantities are given by \cite{Barbosa}
\ba\label{eq2.3d}
&&\Theta(t)={1\over N}\left({\dot{X}\over X}+2{\dot{Y}\over Y}\right)=-{\sqrt{3}\over24}\left(\dot{\beta}+2\dot{\Omega}\right)e^{\sqrt{3}\beta+2\sqrt{3}\Omega}~,\\
&&\sigma(t)={1\over \sqrt{3}N}\left({\dot{X}\over X}-{\dot{Y}\over Y}\right)={1\over24}\left(2\dot{\beta}+\dot{\Omega}\right)e^{\sqrt{3}\beta+2\sqrt{3}\Omega}~, \\
&&l^3(t)=X(t)Y^2(t)=e^{-\sqrt{3}\beta-2\sqrt{3}\Omega}~,
\ea
where $X(t)$ and $Y(t)$ are the same variables as in Eq.(\ref{eq2.0}). As can be seen by their expressions in terms of $X(t)$ and $Y(t)$, these quantities are all diffeomorphism invariant and hence not affected by the choice of the lapse function \cite{Ryan}.

\begin{figure}
\begin{center}
\includegraphics[scale=0.7]{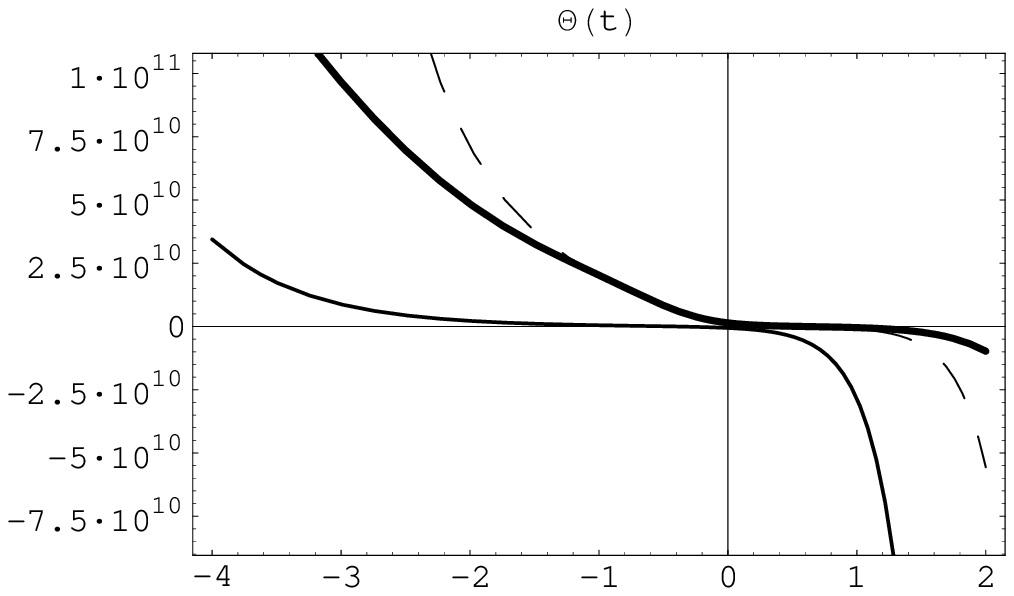}
\includegraphics[scale=0.7]{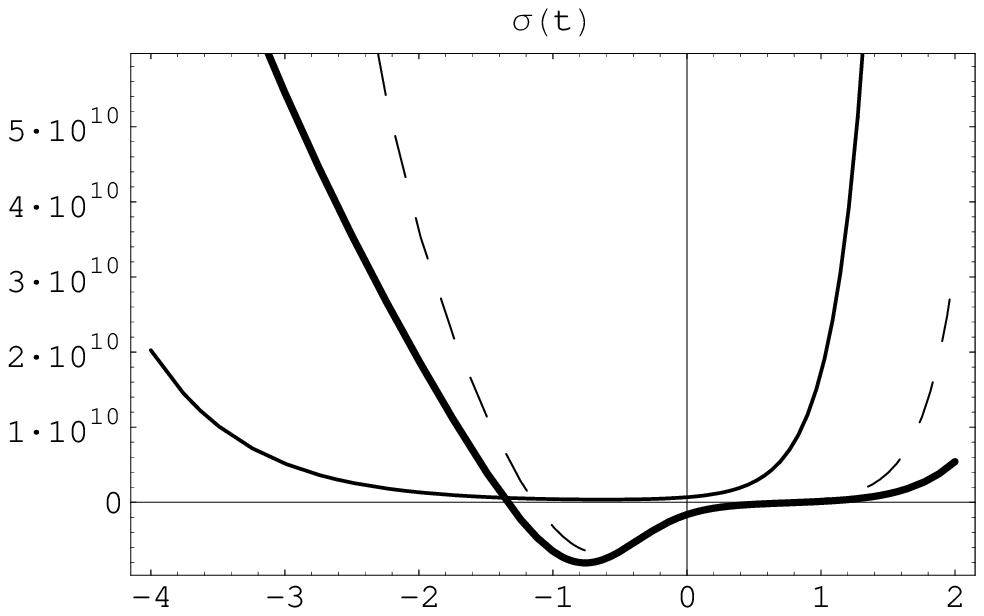}
\includegraphics[scale=0.7]{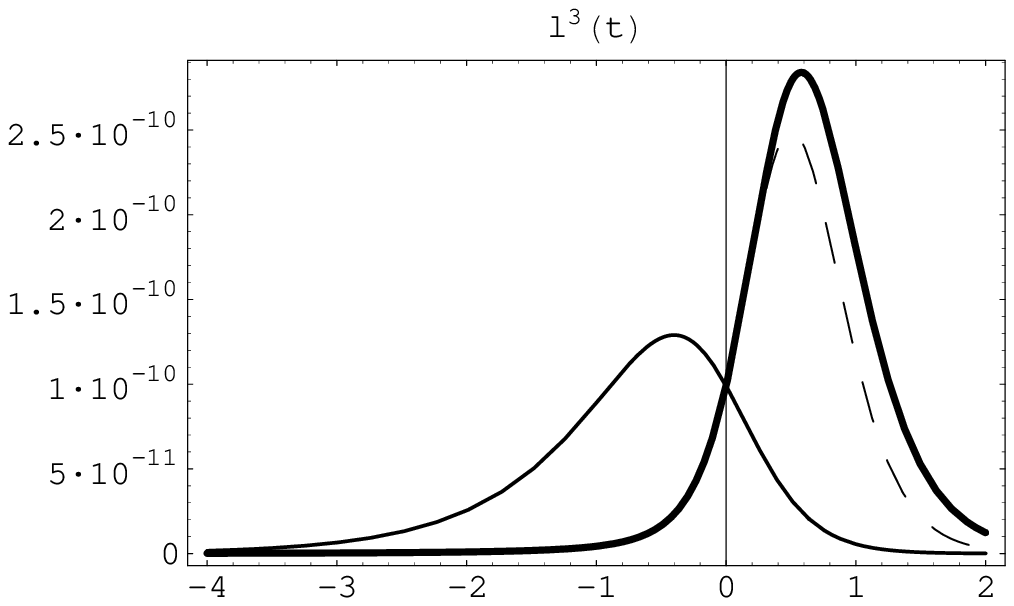}
\caption{Volume expansion, $\Theta(t)$, shear, $\sigma(t)$ and characteristic volume, $l^3(t)$ for (i) $\theta=\eta=0$ (the commutative model) (thin line), (ii) $\theta=5,\eta=0$ (configuration space noncommutative model) (dashed line) and (iii) $\theta=5,\eta=0.1$ (full noncommutative model) (thick line). The initial conditions are $\beta(0)=10$, $P_{\Omega}(0)=0$, $P_{\beta}(0)=0.4$ and $\Omega(0)=1.65$. Notice that the singularity corresponds to $t\rightarrow-\infty$, while the asymptotic region of the metric corresponds to $t\rightarrow\infty$.}
\label{fig:graficosNCQC}
\end{center}
\end{figure}

Through the numerical solutions of the noncommutative classical system Eqs. (\ref{eq2.11}), we are able to obtain estimates for these quantities. In Fig. \ref{fig:graficosNCQC} we depict these results. They are obtained for the commutative KS model and for its extensions displaying noncommutativity in the configuration and in the phase space variables. The thin line exhibits the behavior of the commutative model $(\theta=\eta=0)$. The noncommutative cases are described by dashed $(\theta=5,\eta=0)$ and thick lines for $\theta=5$ and $\eta=0.1$, respectively. Notice that the singularity corresponds to $t\rightarrow-\infty$, while the asymptotic region of the metric corresponds to
$t\rightarrow\infty$.

There are four initial conditions for our problem, $\Omega(0)$, $\beta(0)$, $P_{\Omega}(0)$ and $P_{\beta}(0)$. Three of them, $\Omega(0)$, $P_{\Omega}(0)$ and $P_{\beta}(0)$, are related with each other due to the constraint (\ref{eq2.3a1}). Then, if one chooses numerical values for $P_{\beta}(0)$ and $P_{\Omega}(0)$, one immediately obtains a value for $\Omega(0)$. $\beta(0)$ is an independent initial condition and it is chosen in order to maximize the effect of noncommutativity on the physical quantities.

%%%%%%%%%%%%%%%%%%%%%%%%%%%%%%%%%%%%%%
%%%%%%%%%%%%%%%%%%%%%%%%%%%%%%%%%%%%%%

A simple analysis of the behavior of the physical quantities leads to the conclusion that, for constant values of the initial conditions and $\theta$, the variation of $\eta$ implies that the $\Theta(t)$ and $\sigma(t)$ functions tend towards a straight line for negative $t$ and to zero for positive $t$; on its turn, the overall magnitude of $l^3$ increases with the growth of $\eta$. This is also the pattern for higher values of $\eta$. On the other hand, in the situation where $\eta$ and the initial conditions remain constant and $\theta$ varies, we obtain an analogous qualitative behavior for $\Theta(t)$, $\sigma(t)$ and $l^3(t)$.
Notice that in Fig. \ref{fig:graficosNCQC}, negative $t$ values are considered to depict the complete behavior of the physical quantities. The results are invariant under a time translation, thus the time origin is somewhat arbitrary.

It is relevant to point out that our results are quite stable in what concerns changes of the initial conditions and that the chosen values are fairly typical.
We may however point out that, when $\beta(0)$ assumes negative values, the expansion volume and the shear get smaller, while the characteristic volume gets greater, when compared to the $\beta(0)=0$ case. For $\beta(0)$ positive, the opposite is found. On the other hand, the variation of $P_{\beta}(0)$ has a direct influence on the overall magnitude of the characteristic volume, $l^3(t)$.

The results depicted in Fig. 1 show that from a qualitative point of view, noncommutativity in the configuration variables leads to no noticeable effect when compared with the commutative case in what concerns $\Theta(t)$ and $l^3(t)$; however, a non-trivial noncommutativity makes the behavior of the shear rather ``symmetric" with respect to the arbitrary origin of time, where it assumes the minimal value. The effect is not so sharp for the case of noncommutativity in configuration and momentum variables, but is still clearly present in this situation as well.
Given that the shear corresponds to the distortion in the evolution of the metric, one observes that noncommutativity
implies that the metric is less distorted in its evolution.

Finally, it is important to realize that the analysis of the classical noncommutative model, does not yield any restriction on the possible values of the noncommutative parameters $\eta$ and $\theta$. This feature will actually emerge from the analysis of the quantum version of the noncommutative KS cosmological model.

\subsection{Solutions for the WDW equation}

In this section, all variables are operators. To keep the notation simple, we shall omit the hats on the operators as there is
no risk of confusion here. Let us now consider Eq. (\ref{eq2.9}) in detail. From Eq. (\ref{eq3.3}) we can
define $A=\frac{C}{\sqrt{1-\xi}}$. It then follows that:
\be\label{eq3.4}
\mu P_{\beta_c}+{\eta\over2\mu}\Omega_c=A~.
\ee

As we have already mentioned, the noncommutative WDW equation (\ref{eq2.9}) is fairly complex and does not seem to allow for an analytical solution. Our strategy will consist in solving it numerically by transformation into an ordinary differential equation.
It is easy to verify that the constant of motion Eq. (\ref{eq3.1}) commutes with the Hamiltonian in the constraint space of states, that is
\be\label{eq3.7}
\left[P_{\beta}+\eta\Omega,H\right]=\left[P_{\beta}+\eta\Omega,-P^2_{\Omega}+P^2_{\beta}-48e^{-2\sqrt{3}\Omega}\right]=0~.
\ee
Thus, one can look for solutions of Eq. (\ref{eq2.9}) that are simultaneous eigenstates of the Hamiltonian and of the constraint Eq. (\ref{eq3.4}). If $\psi_a(\Omega_c,\beta_c)$ is an eigenstate of the operator Eq. (\ref{eq3.4}) with eigenvalue $a \in \bkR$, then:
\be\label{eq3.8}
\left(-i\mu{\partial\over\partial\beta_c}+{\eta\over2\mu}\Omega_c\right)\psi_a(\Omega_c,\beta_c)=a \psi_a(\Omega_c,\beta_c)~.
\ee
Solving this equation, one obtains
\be\label{eq3.9}
\psi_a(\Omega_c,\beta_c)=\Re(\Omega_c)\exp{\left[{i\over\mu}\left(a-{\eta\over2\mu}\Omega_c\right)\beta_c\right]}~.
\ee
Substituting the wave function (\ref{eq3.9}) into Eq. (\ref{eq2.9}) yields
\ba\label{eq3.10}
&&{\mu}^2\left[{\Re''\over\Re}-i{\eta\over\mu^2}{\Re'\over\Re}\beta_c-{\eta^2\over4\mu^4}\beta_c^2\right]+i\eta\left[{\Re'\over\Re}-i{\eta\over2\mu^2}\beta_c\right]\beta_c-{\eta^2\over4\mu^2}\beta_c^2+\left(a-{\eta\over2\mu}\Omega_c\right)^2-\nonumber\\
&&-{\eta\over\mu}\left(a-{\eta\over2\mu}\Omega_c\right)\Omega_c+{\eta^2\over4\mu^2}\Omega_c^2-48\exp{\left[-2\sqrt{3}\left(\lambda\Omega_c-{\theta\over2\lambda\mu}\left(a-{\eta\over2\mu}\Omega_c\right)\right)\right]}=0~,
\ea
where $\Re'\equiv\frac{d \Re}{ d \Omega_c}$. After some algebraic manipulations, one gets
\be\label{eq3.11}
\mu^2\Re''+\left(\eta{\Omega_c\over\mu}-a\right)^2\Re-48\exp{\left[-2\sqrt{3}{\Omega_c\over\mu}+{\sqrt{3}\theta\over\lambda\mu}a\right]}\Re=0~.
\ee
Performing the change of variables,
\be\label{eq3.12}
z={\Omega_c\over\mu}\hspace{0.2 cm}\rightarrow\hspace{0.2 cm}{d\over dz}=\mu{d\over d\Omega_c}
\ee
one finally finds for $\phi(z)\equiv\Re(\Omega_c(z))$
\be\label{eq3.13}
\phi''(z)+\left(\eta z-a\right)^2\phi(z)-48\exp{[-2\sqrt{3}z+{\sqrt{3}\theta\over\lambda\mu}a]}\phi(z)=0~.
\ee
This second order ordinary differential equation can be solved numerically. The equation itself depends on the eigenvalue $a$ and on the noncommutative parameters $\theta$ and $\eta$.

%%%%%%%%%.%%%%%%%%%%%%%%%%%%%%%%%%%%%%%
\begin{figure}
\begin{center}
\subfigure[ ~$\theta=\eta=0$ and $a=0.4$]{\includegraphics[scale=0.7]{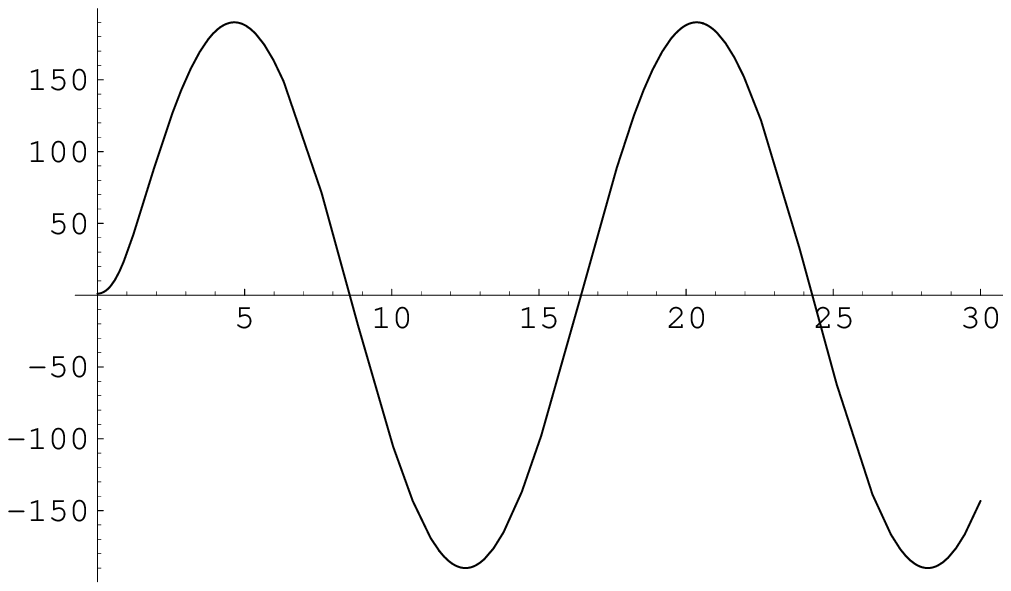}}
\subfigure[ ~$\theta=5$, $\eta=0$ and $a=0.4$]{\includegraphics[scale=0.7]{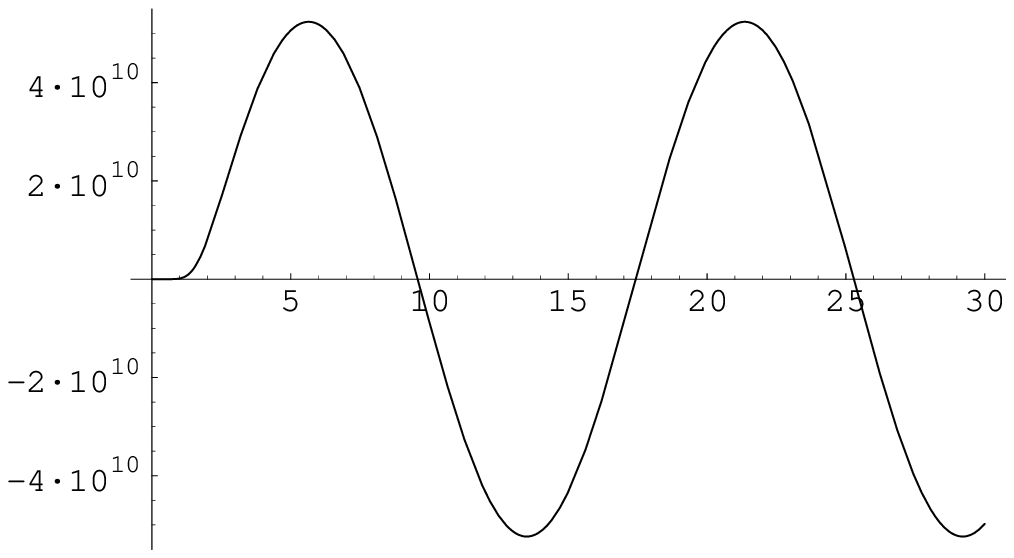}}
\subfigure[ ~$\theta=0$, $\eta=0.1$ and $a=0.565$]{\includegraphics[scale=0.7]{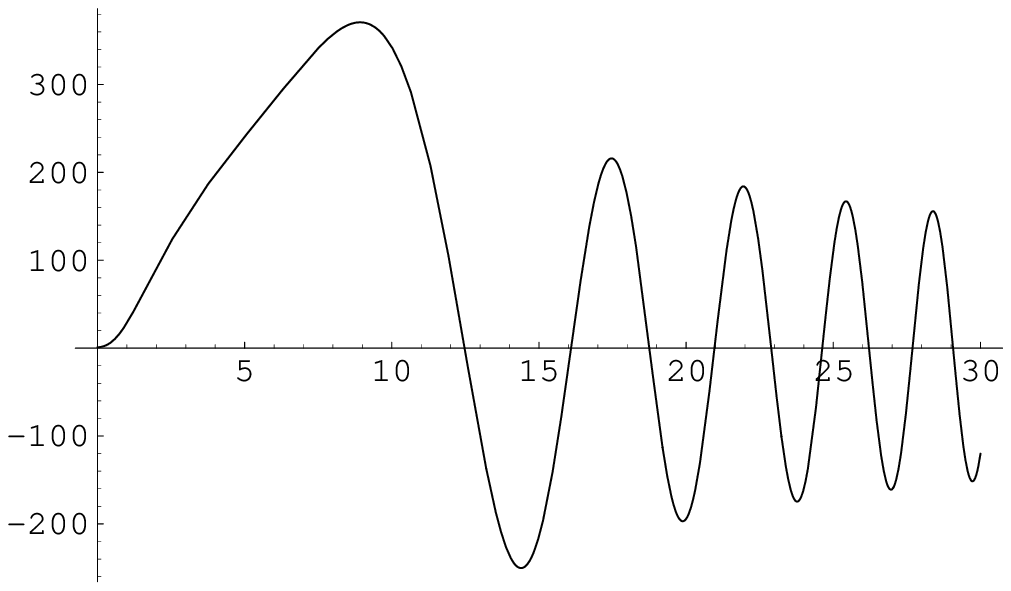}}
\subfigure[ ~$\theta=5$, $\eta=0.1$ and $a=0.799$]{\includegraphics[scale=0.7]{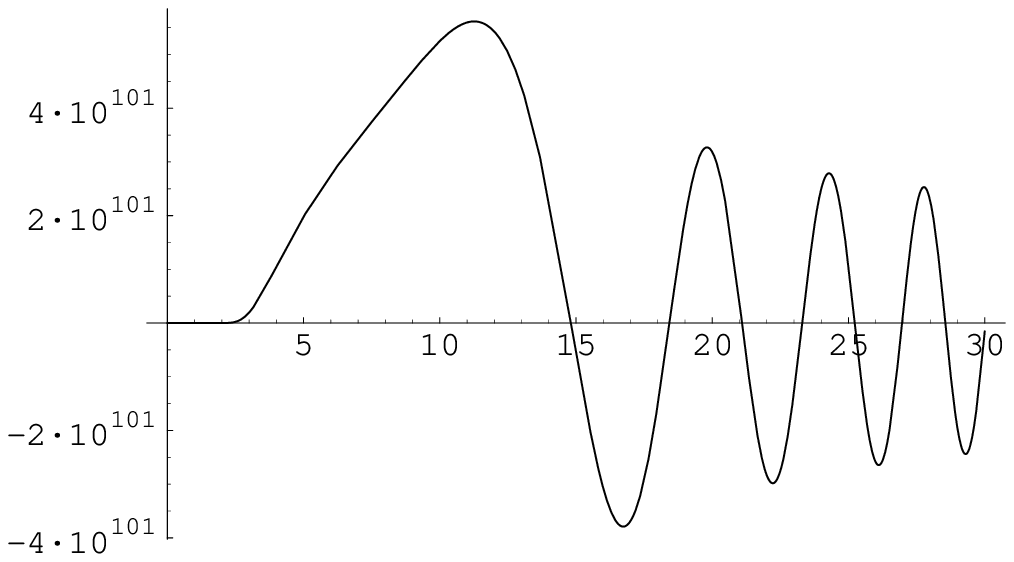}}
\caption{Representation of the numerical solutions of Eq. (\ref{eq3.13}) for different values to the noncommutative parameters. In the four plots $P_{\beta}(0)=0.4$ and $\Omega(0)=1.65$.}
\label{funcaodeonda}
\end{center}
\end{figure}
%%%%%%%%%%%%%%%%%%%%%%%%%%%%%%%%%%%%%%

Fig. \ref{funcaodeonda} depicts numerical solutions of Eq. (\ref{eq3.13}) for particular choices of values of $a, \theta$ and $\eta$.
The noncommutative parameters $\theta$ and $\eta$ are assumed to be in the range of values of the previous classical analysis. The
eigenvalue $a$ was taken to be $a=\frac{C}{\sqrt{1-\theta\eta}}$ and is determined through Eq. (\ref{eq3.1}) from
the classical values $P_{\beta}(0)$ and $\Omega(0)$ used to generate the solutions of Eqs. (\ref{eq2.11}).

The qualitative features of the solutions displayed in Fig. 2 remain within a suitable range of variation of $\theta,\eta$ and $a$. The choice $\theta=5$ is fairly typical in what concerns the properties of the wave function. Furthermore, it is consistent with the point of view that the noncommutative parameters should be of order one close to the fundamental quantum gravity scale. After these general remarks, we are in position to list the most salient features of our results:

\begin{enumerate}
\item For $\theta=5$, we find that the wave function is ill defined (it blows up) for $\eta_c>0.12$, suggesting a system's upper limit for momenta noncommutativity;

\item For fixed $\theta=5$, the variation of $\eta$ yields a wave function with damping behavior for $\eta$ in the range $0.05<\eta<0.12$;

\item The lower limit for $\eta$ having a damping impact on the quantum behavior of the system seems to be around $\eta\sim0.05$ for all $\theta>\eta$. Clearly, higher $\eta$ values (c.f. Fig. \ref{funcaodeonda}) have a great influence in the wave function. Indeed, for $\eta=0$ the wave function simply oscillates. For $0<\eta<0.05$, the wave function is actually amplified instead of exhibiting a damping behavior;

\item The variation of $\theta$ affects the numerical values of $\phi(z)$, but qualitative features of the wave function remain unchanged. For instance, for $\theta=4$ and maintaining the initial conditions, the damping occurs at a slightly different range, $0.07<\eta<0.16$;

\item The qualitative features of the wave function for large $z$ are essentially the ones depicted in Fig. \ref{funcaodeonda} for $z\leq30$;

\item For $\eta>\theta$, the damping behavior of the wave function is more difficult to observe. For instance,
\begin{itemize}
	\item If $\eta=1$, the wave function has a damping behavior for $0<\theta<0.83$. For $\theta>0.83$ it blows up.
\end{itemize}
\begin{itemize}
	\item If $\eta=2$, for $0<\theta<0.1$, the damping behavior is observed, however for $\theta$ values greater than $0.1$, the wave function is ill defined.
\end{itemize}
\begin{itemize}
	\item Finally, if $\eta\geq3$ there are no possible ranges for $\theta$ for which the wave function is well defined.
\end{itemize}
\end{enumerate}

Our criterion to determine bounds for the noncommutative parameters is based on the existence of well defined smooth solutions of the WDW equation. These solutions, as we have seen, do not exist for arbitrary values of $\theta$ and $\eta$. We should point out that if, in addition, we could provide a gauge invariant measure then, at least in principle, we would be able to determine supplementary bounds on $\theta$ and $\eta$ by requiring the formalism to yield finite probabilities. Unfortunately, the issues of gauge fixing and of defining a suitable inner product remain still open problems in quantum cosmology and are, of course, beyond the scope of this paper.

One should notice that it is the momenta noncommutativity that has the strongest impact on the functional form of the eigenstates of the Hamiltonian constraint as we clearly see in Fig. \ref{funcaodeonda}. One of the most interesting features of the introduction of the momenta noncommutativity is that it turned the fundamental solutions of the Hamiltonian constraint from featureless oscillations into damped wave functions displaying an ``almost" normalizable functional form. This is a welcome property as it introduces features in the wave function, selecting the preferred states for the quantum cosmological model and hints on the set of suitable initial conditions for the classical cosmological model. Moreover, this property suggests that for other cosmological models or other types of noncommutativity, the Hamiltonian constraint may display a discrete spectrum and thus normalizable eigenstates. This would be a major breakthrough allowing one to bypass the problem of introducing a measure and to gauge fix in order to obtain finite probabilities. This issue will be further discussed in the next section.

\section{Conclusions}

In this work we have studied the effects of phase space noncommutativity on the minisuperspace KS quantum cosmological model and examined its most distinctive features. We found that despite the difficulty of the problem, a constant of motion could be identified allowing for numerical solutions of the WDW equation in the noncommutative setting.

The resulting solution allows one to study the behavior of the dynamical functions, and classically determine the behavior of the volume expansion, shear and characteristic volume. The evolution of these quantities is obtained for a particular set of initial conditions. We find that the classical effects associated with momenta noncommutativity, $\eta\neq0$, are qualitatively different from those with $\eta=0$, but $\theta\neq0$, in particular in what concerns the shear. Furthermore, for positive and large values of $\beta(0)$, the volume expansion and the shear are quite huge, while the characteristic volume is extremely small.
At the quantum level the effects of noncommutativity are more profound as, on the one hand the existence of solutions for the WDW equation imposes bounds on the noncommutative parameters and, on the other hand, the momenta noncommutativity introduces a damping behavior in the wave function for growing values of the $\Omega$ variable. This implies that the wave function is more peaked for small values of $\Omega$, which is a rather interesting and new portrait of the quantum aspects of the very early universe and is entirely due to the introduction of the momentum noncommutativity.

Notice that some authors choose to convolute the fundamental wave functions with certain kernels (typically Gaussians) to obtain wave functions with features.
For instance in Ref. \cite{Compean}, the wave function for $\eta=0$ has been constructed
\be\label{eq4.4}
\psi (\Omega, \beta) = {\cal N} \int_{- \infty}^{+ \infty} d \nu ~ e^{- a (\nu - b)^2} \psi_{\nu} (\Omega, \beta),
\ee
where
\be\label{eq4.5}
\psi_{\nu} (\Omega, \beta) = e^{i \nu \sqrt{3} \beta} K_{i \nu}
\left( 4 e^{- \sqrt{3} \Omega + \frac{3}{2} \nu \theta} \right),
\ee
and an expression for $\left| \psi (\Omega, \beta) \right|^2$ is depicted for certain values of $a,b$. This wave function also displays a damping behavior. However, this is a consequence of the convolution with the Gaussian and not of the structure of the fundamental solutions $\psi_{\nu} (\Omega, \beta) $ which, in this case, are just featureless oscillations (see Fig. 2 (b)). Indeed the sole effect of the configuration space noncommutativity is a shift in the solution of the commutative model (see Figs. 2(a) and 2(b)). This is in sharp contrast with momenta noncommutativity where, as indicated in Figs. \ref{funcaodeonda}(c) and \ref{funcaodeonda}(d), the damping of the fundamental solutions occurs even in the absence of the configuration space noncommutativity.

We remark nevertheless, that the canonical noncommutativity considered in this work is not by any means unique. There are other admissible deformations of the Heisenberg algebra (\cite{Madore}-\cite{Lorek}). A deformation of the fundamental algebra is tantamount (see Eqs. (\ref{eq2.2}), (\ref{eq2.9})) to adding new interactions to the Hamiltonian. In our case, the inclusion of momentum noncommutativity has led to a damping effect on the wave function of the universe, but not quite enough to render it normalizable. It seems like an interesting avenue to explore whether other types of noncommutativity (as well as other cosmological models) might yield normalizable solutions of the noncommutative WDW equation. A simple example reveals that this is indeed a possibility worth exploring. Even with the simple noncommutative extension considered in this work, one may turn a Hamiltonian constraint with a continuous spectrum (and non-normalizable eigenfunctions) into one with a discrete spectrum (and normalizable eigenfunctions). Consider the non-relativistic, general parametrized system described by a Hamiltonian constraint of the form:
$$
\hat H = \hat P_{\Omega}^2 + \hat P_{\beta}^2 - \delta.
$$
We use the same notation as previously to avoid unnecessary complications. Here $\beta$, $\Omega$ are arbitrary configuration variables and $\delta$ is a positive real constant. The physical states are the solutions of the eigenvalue equation:
$$
\hat H \psi =0
$$
If $\hat P_{\Omega}, \hat P_{\beta}$ commute, then $\hat H$ has a continuous spectrum and the physical wave functions (plane waves) are not normalizable.
The standard procedure to obtain physical predictions for this system is to introduce a measure yielding finite probabilities \cite{Henneaux,Rovelli,Gambini}. We see that, in spite of being a non-relativistic system, this model displays some of the features of our cosmological model.

We now introduce the noncommutativity and set $\left[ \hat P_{\Omega}, \hat P_{\beta} \right] = i \eta$. By performing the SW transformation as in eq.(17) we obtain:
$$
\hat H = {\hat P_{\Omega_c}^2\over2M}+ {\hat P_{\beta_c}^2\over2M} + {1\over2} M \omega^2 \hat{\beta}_c^2 + {1\over2} M \omega^2 \hat{\Omega}_c^2 - \omega \hat L_z -\delta,
$$
where $M = \frac{1}{2 \mu^2}$, $\omega =\eta$ and $\hat L_z = \hat{\Omega}_c \hat P_{\beta_c} - \hat{\beta}_c \hat P_{\Omega_c}$. This is the Hamiltonian of a two dimensional harmonic oscillator coupled to an external constant magnetic field, which is well known in the context of the Landau problem. This Hamiltonian has discrete spectrum and normalizable eigenfunctions \cite{Gamboa}. Hence the physical states, solutions of the constraint equation $\hat H \psi =0$ (for certain values of $\delta$), yield finite probabilities, and we can avoid the problem of introducing a measure. Notice that, as usual, the solutions of the noncommutative constraint $\hat H \psi =0$ are functions of the commutative variables, i.e. they are of the form $\psi(\Omega_c,\beta_c)$. This is the standard procedure. Since $\psi$ is normalizable, probability distributions can then be constructed independently for the noncommutative variables $\Omega$ and $\beta$ (this issue will be further discussed in the next paragraph). In conclusion: our example suggests that, in the context of the quantization of general parametrized systems, finite probabilities can be obtained through (momentum) noncommutativity, at least for non-relativistic finite dimensional models. It remains an open question whether this is also possible in the context of quantum cosmology.

Finally, let us briefly discuss some related issues on the status of the amplitude $\left|\psi (\Omega, \beta) \right|^2$ in the context of noncommutative systems. The quantity $\left|\psi (\Omega, \beta) \right|^2$ cannot be interpreted as a joint probability distribution for $\Omega$ and $\beta$ (even upon smoothing it with a kernel), as these variables do not commute. It is a well known fact that noncommuting variables (such as position and momentum) can at best be statistically described by a quasi-probability distribution (such as the Wigner function \cite{Wigner}), due to Heisenberg's uncertainty principle. The exact expression for such a quasi-probability distribution would be \cite{Bastos1}:
\be\label{eq4.1}
\psi (\Omega, \beta) \star_{\theta} \overline{\psi (\Omega, \beta)},
\ee
for a single noncommutative scale factor, $\Omega$, or $\beta$ $(\eta=0)$. Here $\star_{\theta}$ is the Moyal product \cite{Moyal}:
\be\label{eq4.2}
A(\Omega, \beta) \star_{\theta} B (\Omega , \beta)  = A (\Omega , \beta) \exp \left\{{i \theta\over2} \left({{\buildrel { \leftarrow}\over\partial}\over\partial \Omega} { {\buildrel { \rightarrow}\over\partial}\over\partial \beta} -  {{\buildrel { \leftarrow}\over\partial}\over\partial \beta} {{\buildrel { \rightarrow}\over\partial}\over\partial \Omega}  \right) \right\} B (\Omega , \beta),
\ee
where ${\buildrel { \leftarrow}\over\partial}$ and ${\buildrel { \rightarrow}\over\partial}$ act on $A$ and $B$, respectively.

When the momenta are also noncommutative $(\eta \ne 0)$, it is proven that the quantity Eq. (\ref{eq4.1}) must be replaced by \cite{Bastos2}:
\be\label{eq4.3}
{1\over\varepsilon^2} \psi \left( {\Omega\over\varepsilon}, {\beta\over\varepsilon} \right) \star_{\theta} \overline{\psi \left( {\Omega\over\varepsilon}, {\beta\over\varepsilon} \right)},
\ee
where $\varepsilon$ is the free dimensionless parameter from the SW map (\ref{eq2.8}).

Returning to the work of \cite{Compean}, the correct expression for the amplitude associated to Eq. (22) should actually be given by Eq. (\ref{eq4.1}):
\be\label{eq4.6}
\psi (\Omega, \beta) \star_{\theta} \overline{\psi (\Omega, \beta)} = {\cal N}^2 \int_{- \infty}^{+ \infty} d \nu \int_{- \infty}^{+ \infty} d \mu ~ e^{- a (\nu - b)^2 - a (\mu - b)^2} \left[\psi_{\nu} (\Omega, \beta) \star_{\theta} \overline{\psi_{\mu} (\Omega, \beta)} \right]~.
\ee
A simple calculation using the Bopp shift representation of the $\star_{\theta}$ product
\be\label{eq4.7}
A (\Omega, \beta) \star_{\theta} B (\Omega , \beta) = A \left( \Omega, \beta - {i \theta\over2} {{\buildrel { \rightarrow}\over\partial}\over\partial \Omega}  \right)  B \left(\Omega , \beta + {i\theta\over2}{{\buildrel { \leftarrow}\over\partial}\over\partial \Omega} \right)
\ee
yields:
\be\label{eq4.8}
\psi (\Omega, \beta) \star_{\theta} \overline{\psi (\Omega, \beta)} ={\cal N}^2 \int_{- \infty}^{+ \infty} d \nu \int_{- \infty}^{+ \infty} d \mu ~ e^{- a (\nu - b)^2 - a (\mu - b)^2}  K_{i \nu} \left(4 e^{- \sqrt{3} \Omega + {3\over2} \theta (\nu + \mu) } \right)
\overline{K_{i \mu} \left(4 e^{- \sqrt{3} \Omega - {3\over2} \theta (\nu - \mu) } \right)}~.
\ee
Since one is unable to solve the WDW equation analytically for the $\eta \ne 0$ case, one cannot write down the corresponding expression for Eq. (\ref{eq4.3}). To summarize, noncommutative quantum mechanics is not just ordinary quantum mechanics with additional interactions (via the SW map). This procedure is just an artifact to solve the problem (typically an eigenvalue equation). However, one still has to extract the physical predictions (expectation values, probabilities). The point of noncommutative quantum mechanics is that the physical configuration (or momentum) variables do not commute and one has to resort to expressions such as Eqs. (\ref{eq4.1}) and (\ref{eq4.3}) to make the right predictions.

An interesting aspect of this quasi-probability formulation is that these distributions $\psi \star_{\theta} \overline{\psi}$ may, and usually do, assume negative values, which precludes their interpretation as probability measures. It would be interesting to investigate under which conditions, that is to say for which range of values for $\theta$, $\eta$, will these quasi-distribution be (almost) point wise non-negative, as this would signal the emergence of a commutative universe.

\subsection*{Acknowledgments}

\vspace{0.3cm}

\noindent The work of CB is supported by Funda\c{c}\~{a}o para a Ci\^{e}ncia e a Tecnologia (FCT). The work of OB is partially supported by the FCT project No. POCTI/FP/63916/2005. The work of NCD and JNP was partially supported by Grant No. POCTI/0208/2003 of the FCT.

%\vspace{0.3cm}

\end{document}